\documentclass[conference]{IEEEtran}
\pdfoutput=1
\usepackage{graphicx}
\usepackage{wrapfig}
\usepackage{adjustbox}
\usepackage{subcaption}
\usepackage[utf8]{inputenc}
\usepackage{amsmath}
\usepackage{graphicx}
\usepackage{url}
\usepackage{soul,color}
\usepackage[table]{xcolor}
\usepackage{hyperref}
\usepackage{hyperref}
\hypersetup{
    colorlinks=true,
    linkcolor=blue,
    filecolor=magenta,      
    urlcolor=cyan,
}

\begin{document}

\title{Ultrasound-based Control of Micro-Bubbles for Exosome Delivery in Treating COVID-19 Lung Damage}
\author{
\IEEEauthorblockN{
{\bf Bruna Fonseca},\IEEEauthorrefmark{1}
{\bf Caio Fonseca},\IEEEauthorrefmark{1}
{\bf Michael Barros},\IEEEauthorrefmark{3}
{\bf Mark White},\IEEEauthorrefmark{2}
{\bf Vinay Abhyankar},\IEEEauthorrefmark{4}\\
{\bf David A. Borkholder},\IEEEauthorrefmark{4}
{\bf Sasitharan Balasubramaniam}\IEEEauthorrefmark{1}
}
\\
\IEEEauthorrefmark{1} Telecommunication Software \& Systems Group, Waterford Institute of Technology, Waterford, Ireland \\
\IEEEauthorrefmark{2} Research, Innovation \& Graduate Studies, Waterford Institute of Technology, Waterford, Ireland \\
\IEEEauthorrefmark{3} School of Computer Science and Electronic Engineering, University of Essex, Colchester, UK \\
\IEEEauthorrefmark{4} Kate Gleason College of Engineering, Rochester Institute of Technology, Rochester, USA \\
\\
Emails: bfonseca@tssg.org, cfonseca@tssg.org, m.barros@essex.ac.uk, VPResearch@wit.ie, \\ vvabme@rit.edu, david.borkholder@rit.edu, sasib@tssg.org 
}
\maketitle 

\begin{abstract}
The recent COVID-19 pandemic has resulted in high fatality rates, especially for patients who suffer from underlying health issues. One of the more serious symptoms exhibited from patients suffering from an acute COVID-19 infection is breathing difficulties and shortness of breath, which is largely due to the excessive fluid (cellular leakage and cytokine storm) and mucoid debris that have filled lung alveoli, and reduced the surfactant tension resulting in heavy and stiff lungs. In this paper we propose the use of micro-bubbles filled with exosomes that can be released upon exposure to ultrasound signals as a possible rescue therapy in deteriorating COVID-19 patients. Recent studies have shown that exosomes can be used to repair and treat lung damage for patients who have suffered from the viral infection. We have conducted simulations to show the efficacy of the ultrasound signals that will penetrate through layers of tissues reaching the alveoli that contains the micro-bubbles. Our results have shown that ultrasound signals with low frequencies are required to oscillate and rupture the polymer-based micro-bubbles. Our proposed system can be used for patients who require immediate rescue treatments for lung damage, as well as for recovered patients who may suffer from viral relapse infection, where the micro-bubbles will remain dormant for a temporary therapeutic window until they are exposed to the ultrasound signals.
\end{abstract}


\IEEEpeerreviewmaketitle

\section{Introduction}

The recent COVID-19 pandemic has resulted in many new challenges for humanity in the 21st century. According to the World Health Organization (WHO), more than a 100 million people worldwide have been infected with the SARS-CoV-2 virus, and unfortunately 2.2\% of those infected have resulted in death, this mortality rate was obtained from the number of confirmed deaths divided by the number of confirmed cases from the data made available by the WHO in the time of writing this paper. While the virus does not have a high fatality rate, the risk lies with people who have underlying health condition such as respiratory disorders and chronic diseases, as well as the elderly age group due to their frailty. During 2020, numerous studies have been carried out to gain more knowledge about this unknown virus, and in particular in understanding its infection process, as well as novel treatment techniques.  

The research findings related to the infection process found that people infected with the SARS-CoV-2 virus can progress through four different stages of infection \cite{Ayres2020}. The first stage is described as the incubation period where the virus may not be detected and the patient will not show any types of symptoms. The second stage is the period in which the patient will start showing mild symptoms such as fever, malaise and dry cough, and during this stage the patients can be in the pulmonary phase infection period resulting in pneumonia and pulmonary inflammation. During this stage the virus can be detected through testing \cite{Ayres2020}. The third stage is when the patient starts to exhibit severe symptoms similar to Acute Respiratory Distress syndrome (ARDS) with a high viral load \cite{Ayres2020}. The fourth and final stage is when the patient start to recover from the infection and is highly dependent on their immune system response and the treatments that have been applied.

When the immune system does not react positively against the virus, it starts to proliferate throughout the body, affecting tissues and causing  destruction. This is especially the case for cells that have high concentrations of Angiotensin-Converting Enzyme 2 (ACE2), which can be found in certain cells within the lungs, kidney, intestine, arteries and heart \cite{Hamming2004}. The damage due to the inflammatory process in the lungs can result in life-threatening respiratory disorders and extraordinary therapeutic efforts are required to suppress the inflammation and repair the lungs tissues in a timely manner. For this reason, a number of research challenges have been dedicated towards novel treatments that can suppress the inflammatory process due to the infection.  

Recent studies have shown that the Mesenchymal Stem Cells - Derived Exosomes (MSC-EXOS) presents the same therapeutic benefits of MSCs \cite{ODriscoll2020} and can be used for the treatment of lung damages resulting from the SARS-CoV-2 infection due to their immunomodulatory functions that can be used for organs repair \cite{Jayaramayya2020}. These extracellular vesicles need to be carefully isolated and stored in special mediums during the treatment process, and further studies are required to determine optimal dosage of growth factor production during the treatment. In particular, an outstanding research issue is the ability to dynamically control the release depending on the condition of the damaged tissue.

\begin{figure}
    \centering
    \includegraphics[width = 0.95 \linewidth]{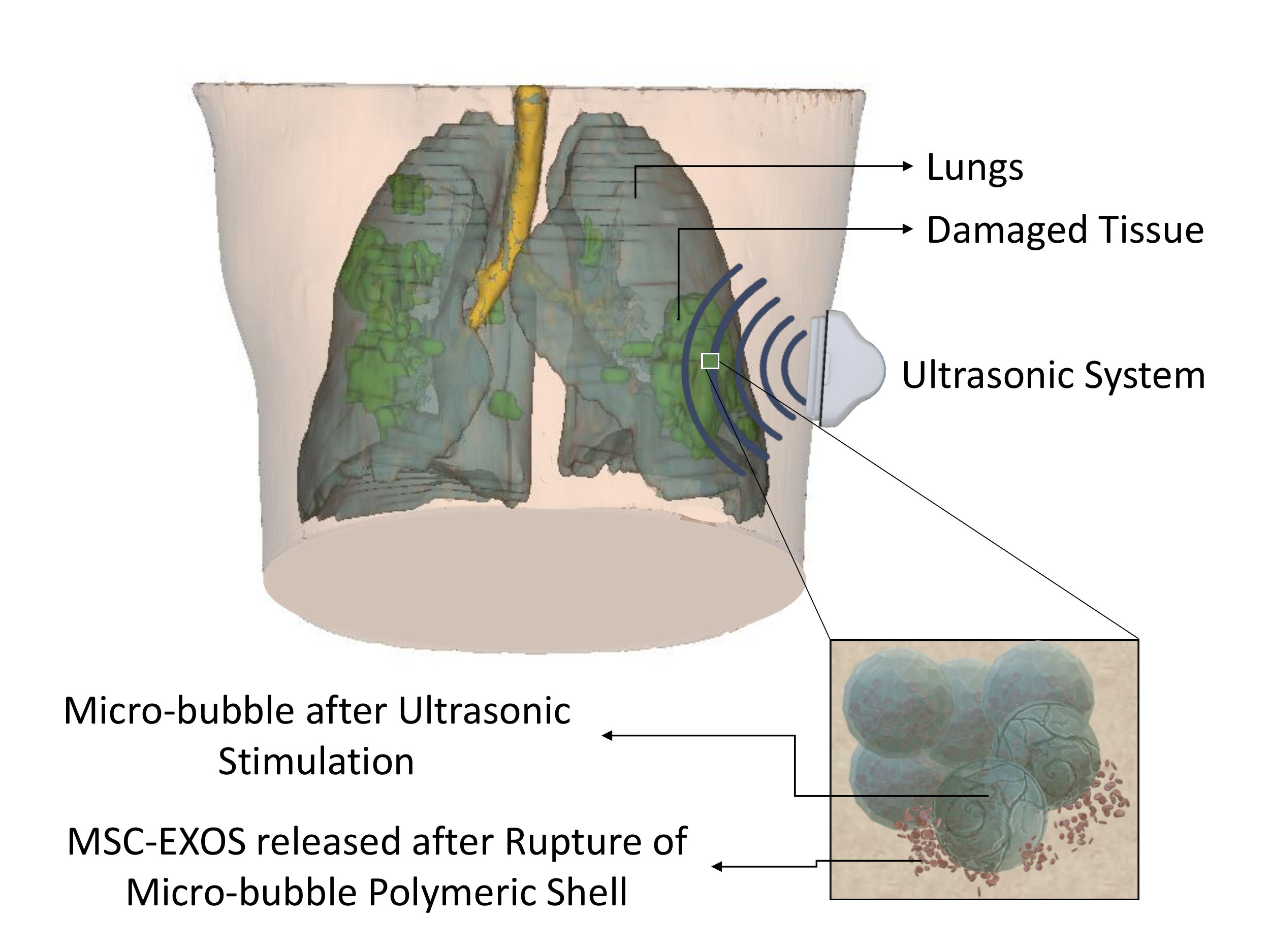}
    \caption{Ultrasound Control of micro-bubble containing exosomes. The ultrasound signals will break the micro-bubbles to release the exosomes.}
    \label{fig:SysArch2}
\end{figure}

This paper proposes a novel solution for an external device that can be used for automated control of MSC-EXOS release to treat patients that are suffering from lung damage due to the COVID-19 infection. This automated mechanism is based on polymer-based encapsulated micro-bubbles that houses MSC-EXOS that are placed in the lungs and remain temporarily on a state of dormancy, where there is a therapeutic window before they are removed by biological actions of the lungs \cite{FernandezTena2012}, until it is activated to be released for treatment. Our proposed solution is illustrated in Fig. \ref{fig:SysArch2}, where the ultrasound source will emit ultrasound waves that will travel deep into the lungs to break the micro-bubbles to release the extracellular vesicles. Our aim is to enable typical wireless body area networks to interface the micro-bubbles within the body that can control functions \cite{Misra2015}\cite{Misra2021}\cite{Misra2015_2}. Besides applying this system to patients who are placed on a ventilator, it can also be used for patients who have recovered from COVID-19, but still remain viral positive with a possibility of infection relapse. In the latter case, the micro-bubbles will remain on a temporary therapeutic window of dormancy in the lungs, and once they are required to release MSC-EXOS, an ultrasound signal can be applied and this could be from a portable device such as mobile phone \cite{butterfly_network}. Simulation has been conducted to demonstrate the concept and in particular on the efficiency of the ultrasound signals that are emitted to break the micro-bubble to release the therapeutic molecules. Our simulation results have found that the ultrasound signal intensity and frequency required is highly dependent on the materials within the alveoli region (e.g., the damage in the lungs that results excessive fluid of mucoid debris (lung edema) caused by the damage). 

The paper is organized as follows: Section II presents the overall system model of the ultrasound that is used to control the breakage of the micro-bubbles. Section III investigates the ultrasound propagation model, considering the attenuation, reflection, and refraction from the different tissue mediums and this impacts on ultrasound intensity. Section IV presents the polymer-based encapsulated micro-bubbles and the models for radius and its impact on a force applied leading to breakage. Section V presents the simulation results, while section VI presents the conclusion. 

\section{Ultrasound Control based Micro-bubble System Architecture}

\begin{figure}
    \centering
    \includegraphics[width = 1 \linewidth]{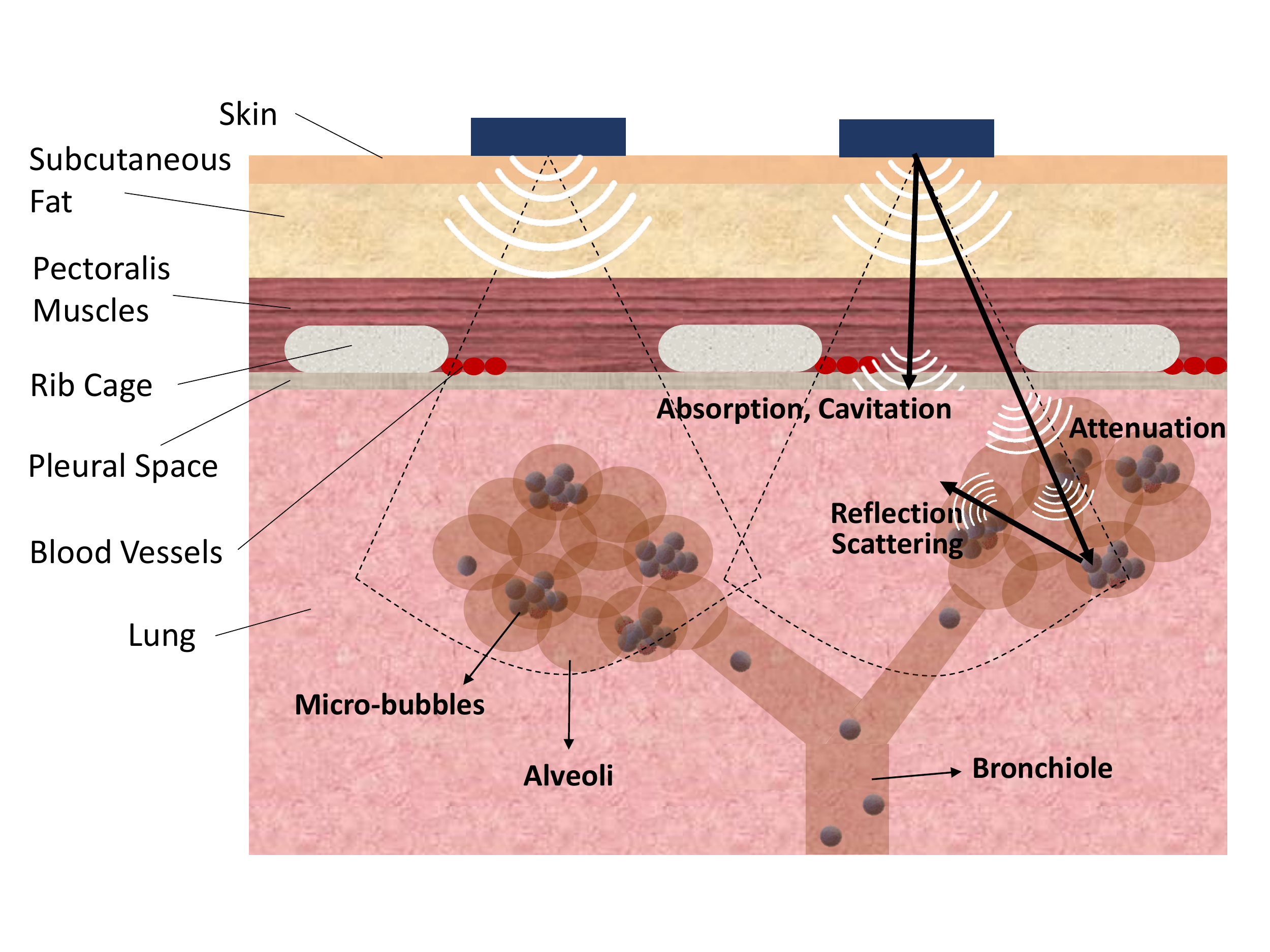}
    \caption{Diagram representing ultrasound propagation through the tissue layers reaching the micro-bubbles within the Alveoli and Bronchiole.}
    \label{fig:UltraS}
\end{figure}

Fig. \ref{fig:UltraS} illustrates our proposed Ultrasound Control based Micro-bubbles system for controlling the release of the exosomes for the tissue repair due to the COVID-19 infection. The  ultrasound transducers are placed externally on the chest of the patients, where it emits ultrasound signals that penetrate through layers of tissue to reach previously placed micro-bubbles. The ultrasound signals will travel through the adipose tissue, pectoralis muscle, pleural space, as well as gaps between the rib cage, and then through the lung tissue to reach the alveoli region where the micro-bubbles are located. Patients can inhale the micro-bubbles containing the exosomes that propagate through the Bronchial tubes reaching and ultimately residing in the vicinity of the alveoli air sacks. The inhalation process can be assisted through the use of the ventilators in endotrachial intubated patients who are in a critical condition. The micro-bubbles are made of polymer material that breaks upon the exposure to sound waves which has been previously observed in \cite{Kooiman2014}\cite{Hoff2000}\cite{Madras2001}\cite{Mayer2008}. 

In this paper, we will focus mainly on two aspects of the overall system: the intra-body ultrasound propagation model as well as the micro-bubbles breaking process. The external transducer that will generate and radiate the ultrasound signals as well as the inhalation process of the micro-bubbles are outside the scope of this paper. Our main objective with the following models is to evaluate and quantify the micro-bubbles breakage process in order to release the exosomes for the treatment. 

\section{Intra-body Ultrasound Propagation Model}

Ultrasound technology has a number of attractive properties that make them ideal for medical applications. This includes their radiation signals that do not cause hazards to the body if they are below certain recommended intensity, and can be integrated into compact low-cost miniature transducer devices that can be easily used. The ultrasound parameters that can be measured such as the signal speed, attenuation, acoustic impedance, and dispersion, can be used and designed appropriately to match different tissue characteristics \cite{Azhari2010}. According to the United States - Food and Drug Administration (US-FDA), the highest known acoustic field emissions $I_{SPTA.3}$ for diagnostic ultrasound devices used for peripheral vessels is $I_{SPTA.3} = 720 mW/cm^2$, for the Cardiac region is $I_{SPTA.3} = 430 mW/cm^2$, for Fetal Imaging \& Other is $I_{SPTA.3} = 94 mW/cm^2$, and for Ophthalmic usage is $I_{SPTA.3} = 17 mW/cm^2$ \cite{FDA2019}. 

The main challenge for our proposed system is to optimize the ultrasound acoustic pressure that is sufficient enough to reach deep into the alveoli region where the micro-bubbles are residing and to break them in order to activate the treatment. We, thereafter, need to consider the ultrasound attenuation caused by the tissue layers that will act as the pathway where the signal will propagate, as well as the acoustic properties of each tissue such as the density, speed of sound and attenuation coefficient. At the same time, we also need to consider other materials that will reside in the alveoli due to the lung damage as this will also affect the signal attenuation.

\subsection{Attenuation}

Ultrasound signals will face attenuation as they propagate through  biological tissue due to the absorbance of energy by the fluids, cells composition and structure of different tissue layers. The attenuation affects higher frequency transmissions, thus resulting in the signal travelling lower distances in the tissue \cite{Azhari2010}. As shown in Fig. \ref{fig:UltraS}, the signals will face a number of effects, and this includes reflection and attenuation as the signals propagate between the different impedance level of each tissue layer.

The attenuation model for the ultrasound signal, which gets attenuated due to the frequency and distance, is represented as follows \cite{Azhari2010}\cite{Prince2015}: 

\begin{equation} \label{eq:UltraAtt}
I_{d} = I_{s} 10^{-(\frac{\alpha f d}{10})} ,
\end{equation}

\noindent where $I_{d}$ and $I_{s}$ are the ultrasound intensity levels at a distance $d$ from a  source $s$, respectively, $\alpha$ is the attenuation coefficient of the various tissue type, $f$ is
the ultrasound wave frequency, and $d$ is the distance between the source and the target micro-bubbles.

For the micro-bubbles radius oscillation model developed by Hoff, which will be explained in section IV in this article, he used the acoustic pressure as excitation for the micro-bubbles instead of the ultrasound intensity, and this pressure can be expressed as Eq. (\ref{eq:Intensity_Pressure}) \cite{Prince2015}, where $P_d$ is the pressure at a distance $d$ from a source $s$, $I_{d}$ is the intensity at the same distance and $Z_{medium}$ is the acoustic impedance of the medium where the micro-bubble is inserted which can be calculated from the product between the density ($\rho$) and  the speed of sound ($c$) of the medium.

\begin{equation} \label{eq:Intensity_Pressure}
P_{d} = (I_{d}Z_{medium})^{\frac{1}{2}} ,
\end{equation}

\noindent Both Eq. (\ref{eq:UltraAtt}) and Eq. (\ref{eq:Intensity_Pressure}) are going to be used for the simulation process of the ultrasound intensity attenuation through the tissue layers and to determine the acoustic pressure reaching the micro-bubbles at the specified distance within the lungs respectively.

\subsection{Reflection}

As the ultrasound signals penetrate through the different layers of the tissue, it will encounter different acoustic impedance, which can contribute to the reflection of the signals as well. Based on two materials $Z_1$ and $Z_2$ with incident ultrasound signal $I_s$ and the received signal $I_r$ at distance $d$, we can represent this relationship as \cite{Prince2015}

\begin{equation}\label{eq:reflection}
\frac{I_r}{I_s}=\frac{(Z_2 - Z_1)^2}{(Z_2+Z_1)^2}.
\end{equation}

However, we are interested in the transmitted ultrasound signal $I_t$, that will continue propagating through the different tissue layers eventually reaching the micro-bubbles, and it can be obtained from Eq. (\ref{eq:reflection}) as follows \cite{Prince2015}:

\begin{equation}\label{eq:transmission}
I_t = 1 - \frac{I_r}{I_s} = \frac{4Z_2Z_1}{(Z_2+Z_1)^2}.
\end{equation}

\section{Polymer-Based Micro-bubbles for Exosomes Encapsulation}

\begin{figure}
    \centering
    \includegraphics[width = 0.6 \linewidth]{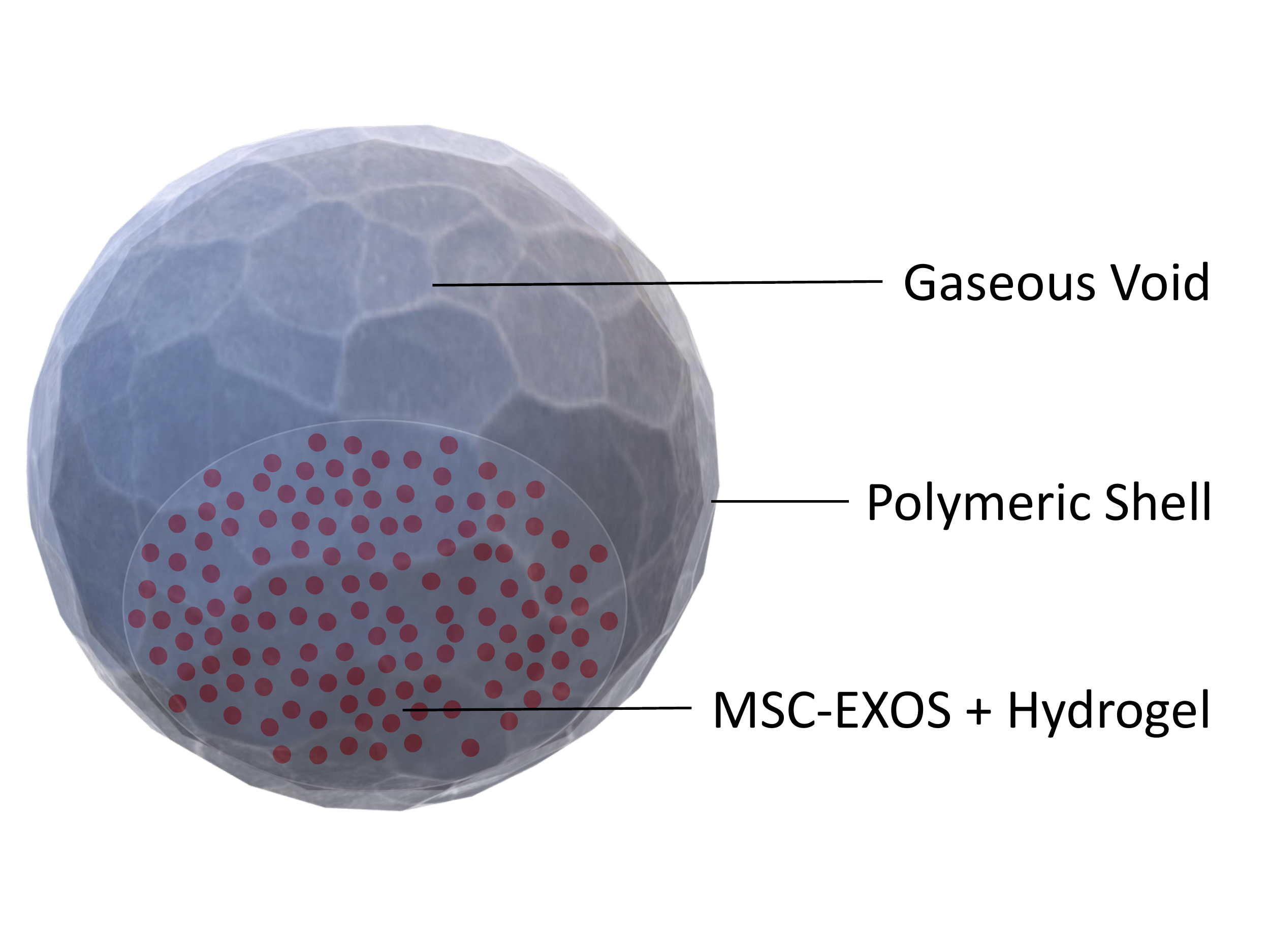}
    \caption{Polymer-based micro-bubble for MSC-EXOS encapsulation with a gaseous nucleus.}
    \label{fig:micro-bubble}
\end{figure}

Over the years numerous studies have been carried out to incorporate drugs and therapeutic molecules within the encapsulated micro-bubbles to help with different treatments that require targeted drug-delivery  \cite{Upadhyay2019} \cite{Kooiman2014} \cite{Bouakaz2016}. For example, micro-bubbles with 0.1$\mu m$ to 10$\mu m$  diameter, have been used  as ultrasound contrast agents (UCAs) to improve the quality of the images that use ultrasound technologies. The most common approach used for micro-bubble based drug-delivery is by either  attaching or inserting the substances into the encapsulating shells (this could be in a two-coated shell). However, our application requires that we rupture and break the micro-bubble, which will have the shell's thickness as one of the main factors to overcome, and so adding another layer, a viscoelastic layer for example, to the shells will limit the micro-bubble's oscillation amplitude, which makes the rupture process very difficult. In particular, when we focus on the release of the exosomes at low acoustic amplitude  \cite{Kotopoulis2015}. Therefore, an alternative solution is to insert the exosomes, contained in the hydrogel fluid, into the gaseous nucleus of the micro-bubble, as illustrated in Fig. \ref{fig:micro-bubble}, which in turn will transform it into an \emph{anti-bubble}.

According to Kotopoulies et al. \cite{Kotopoulis2015}, the dynamics of the anti-bubble is similar to the dynamics of a gas bubble if the liquid core is less than 50\% of the micro-bubble's diameter \cite{Kotopoulis2015}. This way, we can use a simplified version of Church's model \cite{Church1995} that was developed by Hoff et al. \cite{Hoff2000} for the oscillations of polymeric micro-bubbles considering the effects of structure that utilizes an encapsulating shell. The oscillation behaviour of the micro-bubble is represented as:

\begin{align}\label{eq:hoff}
\begin{split}
   \rho_L \left( \ddot{R_2} R_2 + \frac{3}{2}\dot{R_2}^2 \right) = p_{ge}\left( \frac{R_{1e}}{R_1} \right)^{3\kappa} - p_\infty(t) - 4\mu_L\frac{\dot{R_2}}{R_2} \\ - 12\mu_S\frac{{R_{1e}}^2 d_{Se}}{R_2^3}\frac{\dot{R_1}}{R_1} - 12 G_s \frac{{R_{1e}}^2 d_{Se}}{R_2^3} \left( 1 - \frac{R_{1e}}{R_1} \right),
\end{split}
\end{align}

where $R_1$ and $R_2$ are the inner and outer shell radii, $\rho_L$ is the density of the surrounding liquid, $p_{ge}$ is the gas equilibrium pressure within the micro-bubble, $R_{1e}$ and $R_{2e}$ are the inner and outer shell radii at equilibrium, $d_{Se}$ is the shell thickness at equilibrium, $p_\infty(t)$ is the pressure in the liquid far from the micro-bubble, $\kappa$ is the gas polythropic exponent, $\mu_L$ is the shear viscosity of the surrounding liquid, and $\mu_S$ and  $G_s$ are the shear viscosity and shear modulus, respectively, of the polymer micro-bubble shell \cite{Hoff2000}. 

The Eq. (\ref{eq:hoff}) is represented as a function of the inner and outer radii of the micro-bubble. However, Hoff modified Eq. (\ref{eq:hoff}) into a function of the outer radius $R = R_2(t)$. To do that, he used the following expression \cite{Hoff2000}:

\begin{align}\label{eq:hoffsimplification}
\begin{split}
   \frac{R_{1e}}{R_1} \approx \frac{R_{2e}}{R_2}\left( 1 + \left( \frac{d_{Se}}{R_{2e}} - \frac{d_S}{R_2} \right) \right) \approx \frac{R_{2e}}{R_2} = \frac{R_e}{R}
\end{split}
\end{align}

\noindent Hoff also considered that at equilibrium, the pressure in the gas inside the bubble can be assumed to be equal to the hydrostatic pressure of the surrounding liquid ($p_{ge} = p_0$) \cite{Hoff2000}, which means that there is no tension in the shell at equilibrium. In addition, he also considered that the pressure far from the bubble ($p_\infty$) is the sum of the surrounding liquid pressure $p_0$ and the acoustic pressure that is being applied $p_i(t)$ \cite{Hoff2000}.
These assumptions resulted in Eq. (\ref{eq:hofffinal}), which represents the micro-bubble motion equation, described as follows:

\begin{align}\label{eq:hofffinal}
\begin{split}
   \rho_L \left( \ddot{R} R + \frac{3}{2}\dot{R}^2 \right) = p_0\left( \frac{R_e}{R} \right)^{3\kappa} - p_i(t) - 4\mu_L\frac{\dot{R}}{R} \\ - 12\mu_S\frac{{R_e}^2 d_{Se}}{R^3}\frac{\dot{R}}{R} - 12 G_s \frac{{R_e}^2 d_{Se}}{R^3} \left( 1 - \frac{R_e}{R} \right)
\end{split}
\end{align}

\begin{table*}[h!]
    \centering
    \caption{Acoustic Properties of Different Tissues and Materials used in the Simulations.}
    \begin{adjustbox}{width=\textwidth}
    \begin{tabular}{|c|c|c|c|c|}
      
      \hline
      Tissue/Material & Density $(kg/m^3)$ & Speed of Sound $(m/s)$ & Acoustic Impedance $x10^6(kg/sm^2)$ &  Attenuation $(dB/cm/MHz)$
      \\
      \hline
      Skin \cite{Mast2000} & 1090 & 1615 & 1.76 & 0.35 
      \\
      \hline
      Soft Tissue \cite{Mast2000} & 950 & 1478 & 1.404 & 0.48 
      \\
      \hline
      Connective Tissue \cite{Mast2000} & 1120 & 1613 & 1.806 & 1.57 
      \\
      \hline
      Muscle \cite{Mast2000} & 1050 & 1575 & 1.654 & 1.09 
      \\
      \hline
      Healthy Lungs \cite{Shi2021} & 180 & 650 \cite{EDELMAN1988} & 0.177 & 5.66
      \\
      \hline
      COVID-19 fluids and debris \cite{Shi2021} & 220 & 650 \cite{EDELMAN1988} & 0.143 & 4.93
      \\
      \hline
      PLGA \cite{Parker2010} & 1190 & 23260 & 2.76 & 0.42
      \\
      \hline
      Water \cite{Azhari2010} & 1000 & 1480 & 1.48 & 0.002
      \\
      \hline
    \end{tabular}
    \end{adjustbox}
    \label{tab:table2}
\end{table*}

\section{Simulations}
\subsection{Simulation Model}
We developed a simulation model using MATLAB, where we simulate the ultrasound signal intensity that propagates through various layers of the tissue between the chest and the alveoli, and this includes the materials within the alveoli that resulted from the lung tissue damage, as well as signal that penetrates through the micro-bubbles. In our simulation study, we consider five different layers of tissues, and this includes the skin with an average thickness of $2.05mm$ \cite{Derraik2014}, the subcutaneous fat tissue with a thickness of $19.57 mm$ \cite{Derraik2014},  muscle tissue with a thickness of $3.29mm$ \cite{Yoshida2019}, the connective tissue with $0.5mm$, and finally the lung tissue with $10cm$ thickness \cite{Kramer2012}. The acoustic properties of each tissue are presented in Table \ref{tab:table2}. We know that for patients who suffer from severe COVID-19 infections, the lung characteristics changes, where the density is larger when compared to a healthy lung. This expansion is due to the amount of inflammation and mucus produced during the infection and this will depend on the stage as well as the severity of the infection. In cases where the patients suffer from respiratory distress, this expansion can increase considerably. 

\subsection{Ultrasound Propagation through COVID-19 Damaged Tissue}

\begin{figure}
    \centering
    \includegraphics[width = 1 \linewidth]{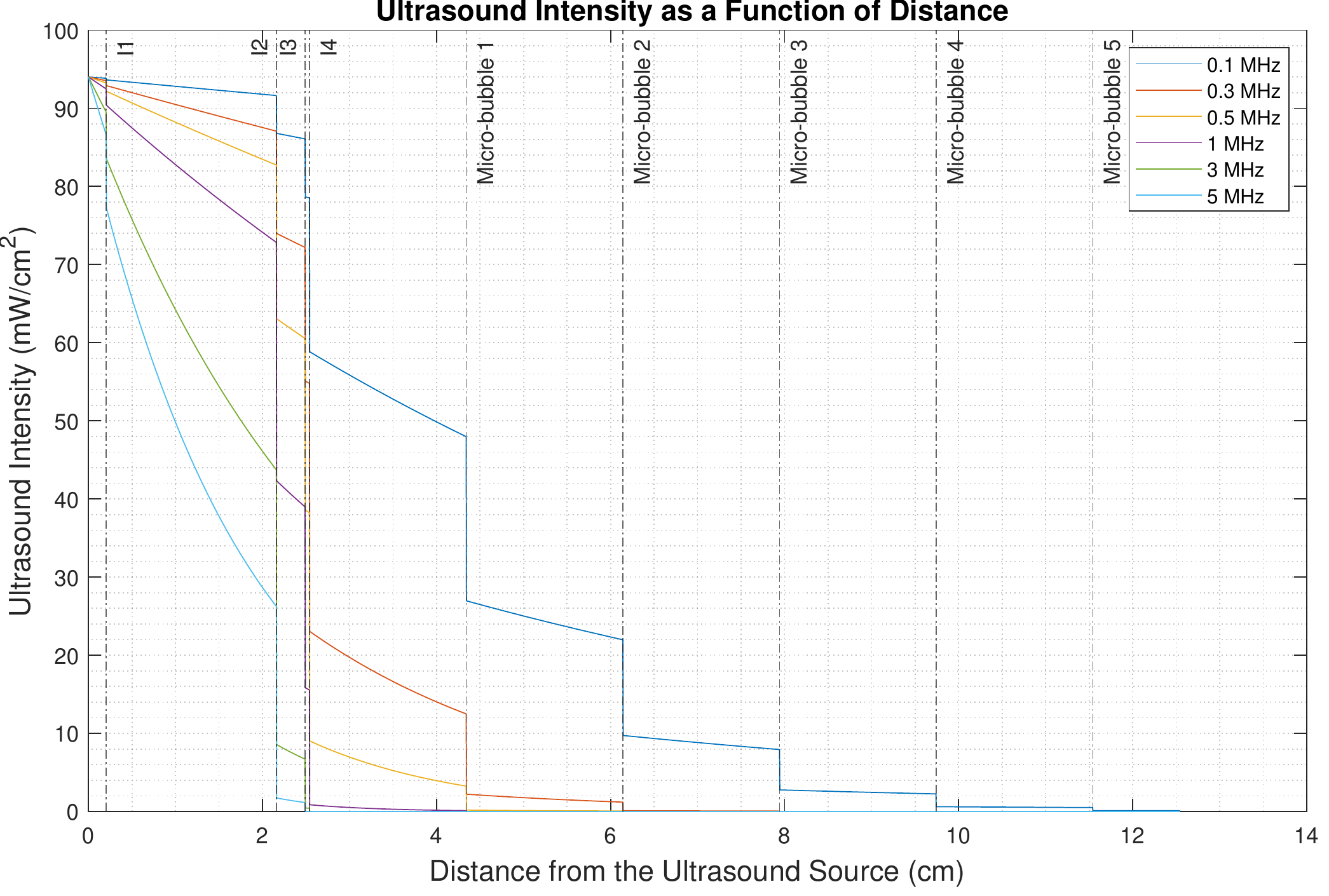}
    \caption{Ultrasound attenuation through all the tissue layers. I1 represents the interface between the skin and soft tissue. I2 is the interface between soft tissue and muscles. I3 represents the interface between muscles and connective tissues. I4 represents the interface between connective tissues to the Lung. The position of the micro-bubbles is also represented in the figure. Variation of frequency from 0.1 MHz to 5 MHz.}
    \label{fig:Attenuations Tissues}
\end{figure}

To run the simulations we developed an algorithm on MATLAB using Eq.(\ref{eq:UltraAtt}) - Eq.(\ref{eq:transmission}) and the results are shown on Fig. \ref{fig:Attenuations Tissues}. In the simulation results presented in Fig. \ref{fig:Attenuations Tissues}, we tested different ultrasound frequencies to see how it penetrates through varying distances. The simulation also includes materials from the COVID-19 patients in the alveoli. The results show that low frequencies are able to penetrate towards the alveoli compared to higher frequencies, which suffer from very high attenuation with distance. All tissue layers caused ultrasound intensity attenuation and reflection, however the layers before the Lung (from the source to $2.541cm$ on Fig. \ref{fig:Attenuations Tissues}) presented a lower reflection and the ultrasound signal did propagate through all of them up to the lungs. When the signal reached the lungs (after $2.541cm$ of distance), it suffered higher absorption and reflection for all the applied frequencies.

\subsection{Micro-bubble Radius Oscillation Simulations}

\begin{figure}
    \centering
    \includegraphics[width = 1.01 \linewidth]{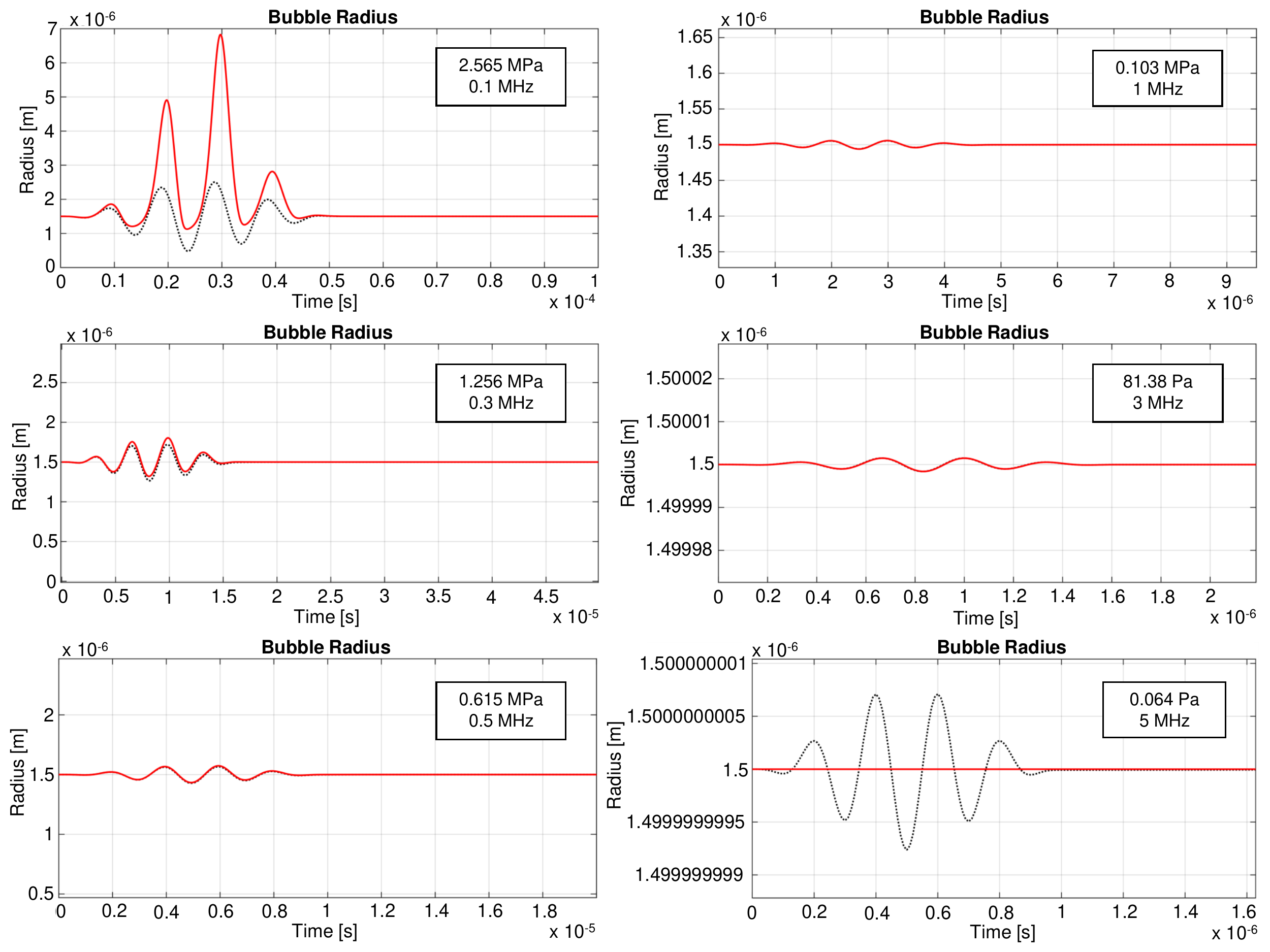}
    \caption{Simulation of micro-bubble excitation with respect to varying ultrasound frequency and intensity. This simulation is for a single micro-bubble. The dashed lines represents the ideal values and the red lines represents the simulated values.}
    \label{fig:Radius OScillation}
\end{figure}

To perform the simulations of the micro-bubbles radius oscillations, the MATLAB "Bubblesim" package developed by Hoff was used. For our study, we considered a micro-bubble with a polymeric shell of poly(lactic-co-glycolic acid) (PLGA), which has a \emph{Shear Modulus} of $60MPa$, and \emph{Shear Viscosity} of $1 Pa-s$ \cite{Coulouvrat2012} with $5 nm$ of thickness and $1.5 \mu m$ for the outer radius. For the first simulations, we considered the gas inside the micro-bubble to have the same properties of the air, and that the micro-bubbles were inserted into the Lung with the medium characteristics of COVID-19 Lung tissue. Fig. \ref{fig:Radius OScillation} shows the amplitude of the oscillations with respect to variations in the ultrasound frequency and pressure for a single micro-bubble. The results comply with ultrasound propagation behaviour in Fig. \ref{fig:Attenuations Tissues}, where we find that at the very low frequency of $0.1MHz$ we will obtain high radius oscillations of the micro-bubbles. At this particular frequency, the amplitude is large enough to rupture the micro-bubble. While at $0.3MHz$ we observe small oscillations, that is not sufficient to break the micro-bubble.

\begin{figure}
    \centering
    \includegraphics[width = 1 \linewidth]{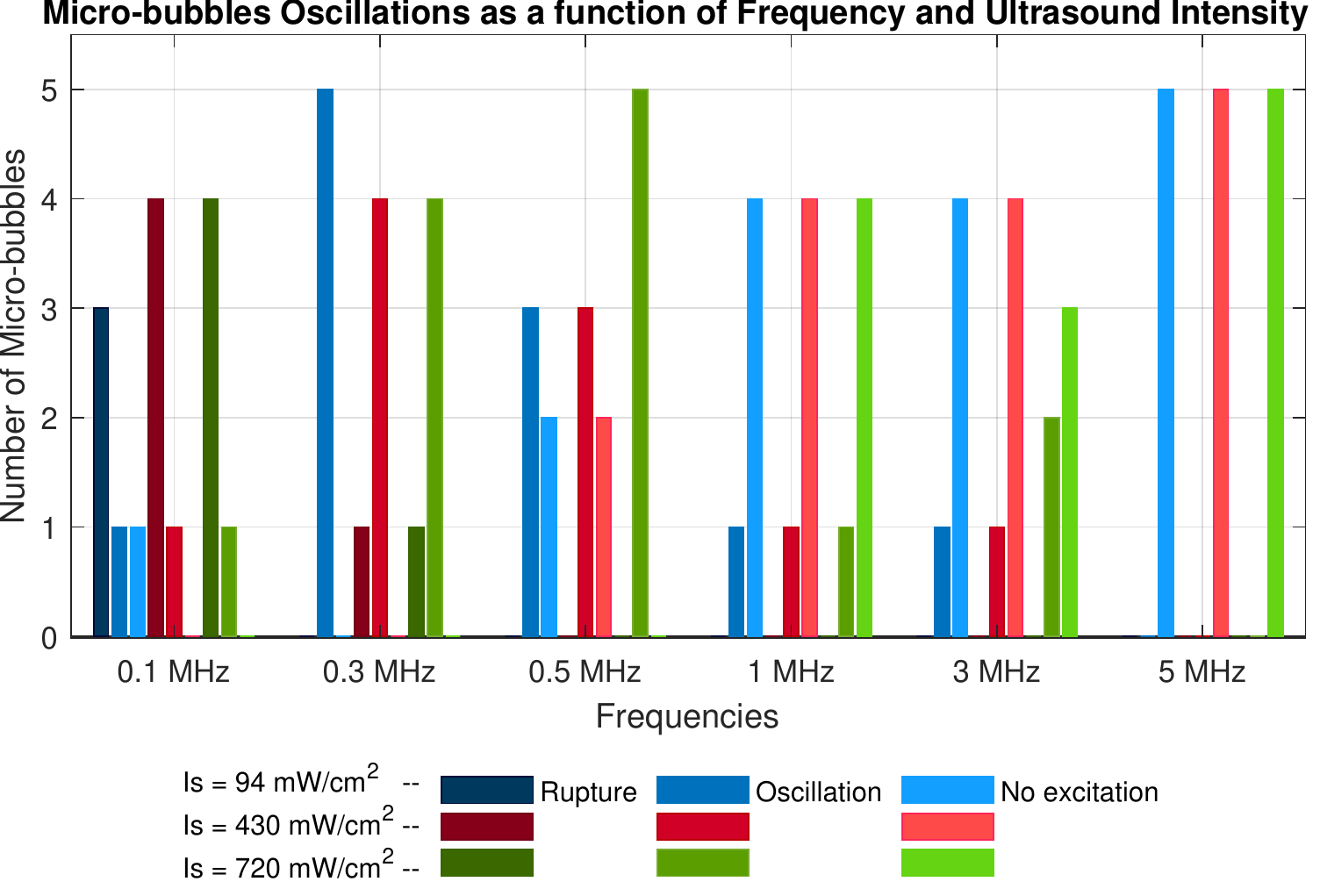}
    \caption{Number of micro-bubbles that ruptured, oscillated or didn't suffer any excitation when exposed with different frequencies and ultrasound intensities.}
    \label{fig:5Microbubbles}
\end{figure}

The same simulation process was done for the other ultrasound intensity limits specified by the US-FDA ($94 mW/cm^2$, $430 mW/cm^2$ and $720 mW/cm^2$) to detect the numbers of micro-bubbles that will break within the lung. The simulation considered five micro-bubbles inside a lung infected with COVID-19, from which the reference values can be observed in Table \ref{tab:table2}. The micro-bubbles are placed at a distance of $1.8cm$ from each other at the same depth. The results from the simulations are shown in Fig. \ref{fig:5Microbubbles}. For the intensity source of $94 mW/cm^2$, three micro-bubbles ruptured and two oscillated with the applied frequency of $0.1MHz$ and for the other frequencies, part of the micro-bubbles oscillated or did not suffer any excitation. When the micro-bubbles were applied $430mW/cm^2$, four micro-bubbles ruptured and one oscillated with $0.1MHz$, one micro-bubble ruptured and the rest oscillated with $0.3MHz$. At the same time some oscillations were observed and for other micro-bubbles no excitation was observed for the other frequencies. For the ultrasound intensity of $720mW/cm^2$, the micro bubbles showed the same results obtained for $430mW/cm^2$ with frequencies of $0.1MHz$ and $0.3MHz$, and for the other frequencies some micro-bubbles oscillated and some did not suffer excitation.

\section{Conclusion}

The COVID-19 pandemic has challenged researchers from many disciplines to develop new and novel treatment solutions in order to rescue patients who are infected and seriously compromised with the virus. In this paper, we proposed an ultrasound-based control of exosomes release from micro-bubbles for treating lung damage from COVID-19 infection. Simulations have been conducted to validate the required intensity and frequency of ultrasound signals that will penetrate different layers of the tissue to reach the alveoli containing the micro-bubbles. The results from the simulations showed that very low frequency signals are more efficient in oscillating and vibrating the micro-bubbles in order to break them to release the exosomes. Our proposed approach can lead to future reactive and proactive treatments, where micro-bubbles can be ruptured to release the exosomes on demand. 

\section*{Acknowledgment}

The research emanated from this publication is funded by the Waterford Institute of Technology Postgraduate Scholarship.

\ifCLASSOPTIONcaptionsoff
  \newpage
\fi

\bibliographystyle{IEEEtran}
\bibliography{references}

\end{document}